\documentclass[prd,superscriptaddress,nofootinbib,showpacs,preprint]{revtex4}

\usepackage{graphicx}
\usepackage[usenames,dvipsnames]{color}
\usepackage{amsmath,amssymb}
\usepackage[colorlinks]{hyperref}

\newcommand{\be}{\begin{equation}}                                             
\newcommand{\ee}{\end{equation}}

\usepackage{epsfig}

\begin{document}
\title{Supersymmetric Left-Right models with Gauge Coupling Unification and Fermion Mass Universality}
\author{Debasish Borah}
\email{debasish@phy.iitb.ac.in}
\affiliation{Department of Physics, Indian Institute of Technology, 
Bombay, Mumbai - 400076, India}
\author{Urjit A. Yajnik}
\email{yajnik@iitb.ac.in}
\affiliation{Department of Physics, Indian Institute of Technology, 
Bombay, Mumbai - 400076, India}
\affiliation{Department of Physics, McGill University, 3600 rue University, Montreal QC H3X 2T8, Canada}
\begin{abstract}
We explore the unification of gauge couplings and fermion masses in two different types of 
supersymmetric left-right models, one with Higgs triplets and the other with both Higgs 
triplets as well as bitriplets. The minimal versions of these models do not give rise to 
the desired unification and some extra fields have to be added. After such a modification,
it is possible in one model to get gauged $B-L$ symmetry to be unbroken down
to TeV scale. We also identify the parameter 
space at the electroweak scale which gives rise to fermion mass unification at a high scale $M_{G}$. 
Type I seesaw emerges as the natural explanation of the small neutrino masses in both the models.
\end{abstract}

\maketitle

\section{Introduction}
The Left-Right symmetric model \cite{Pati:1974yy,Mohapatra:1974gc, Senjanovic:1975rk, 
Mohapatra:1980qe, Deshpande:1990ip} has always been an appealing extension of the 
Standard Model  of particle physics. In such models parity is spontaneously 
broken and the smallness of neutrino masses \cite{Fukuda:2001nk, Ahmad:2002jz, 
Ahmad:2002ka, Bahcall:2004mz} arises in a natural way via seesaw mechanism 
\cite{Minkowski:1977sc, GellMann:1980vs, Yanagida:1979as, Mohapatra:1979ia}. 
Finally $B-L$ number becomes an abelian gauge charge, which has important simplifying
implications for the observed baryon asymmetry of the Universe. 
Anticipating the embedding of this model in an $SO(10)$ unified theory it is plausible 
to assume presence of TeV scale supersymmetry in order to stablize the hierarchy
between the electroweak scale and the unification scale. In the class of
models to be considered here, generically called Supersymmetric Left-Right (SUSYLR) models, the effective potential 
of the spontaneously broken theory permits that the $U(1)_{B-L}$ remains unbroken 
upto low energies, close to TeV scale.
This is the possibility we shall assume with a hope of unearthing both supersymmetry 
and $B-L$ symmetry within collider energy regimes.
In this paper we study whether gauge coupling unification remains viable under 
such conditions, along with a consistent see-saw explanaton for the fermion masses.  
Within the specific models presented here, we achieve partial success of these goals.

\indent 
When the non-supersymmetric model of \cite{Pati:1974yy,Mohapatra:1974gc, Senjanovic:1975rk, 
Mohapatra:1980qe} is extended to incorporate supersymmetry it is found that the effective
potential of the proposed minimal 
model fails to provide spontaneous parity breaking \cite{Kuchimanchi:1993jg}.
One possible direction for ameliaorating this problem is the inclusion of
non-renormalizable terms \cite{Kuchimanchi:1993jg}\cite{Aulakh:1998nn}, or non-perturbative  
corrections from an additional singlet \cite{Babu:2008ep}. 
In an alternative approach, it was proposed in ~\cite{Aulakh:1998nn, Aulakh:1997ba} that addition of
heavy scalars similar to the $\Delta$ triplets of the non-supersymmetric version, 
but neutral under $B-L$, avoid the above stated problems and provide spontaneous parity
breaking with only renormalizable terms considered.
Recently, the spontaneous parity violation was also demonstrated in an alternative supersymmetric 
Left-Right model \cite{Patra:2009wc}  where the extra fields added to the minimal field content 
are a gauge singlet and a bitriplet under $SU(2)_L \times SU(2)_R$. 

These studies remain ad hoc unless additional guiding principles can be used to
reduce the variety of models or indeed to single out any one model. In the present work we
shall seek input from the requirements of gauge coupling unification and fermion mass
universality. However there are additional guiding principles articulated in 
\cite{Aulakh:1999cd, Aulakh:1999pz, Aulakh:2000sn}. One concerns the mass scales of the scalars, 
wherein supersymmetry can give rise to accidentally light particles, dubbed ``survival of the fittest'' 
phenomenon. The fact that the renormalizable superpotential forbids certain categories of terms in
the scalar potential which would have been otherwise permitted by gauge invariance,
ensures that certain scalar masses do not receive large corrections from heavier particles
in the spectrum. In accordance with this, we shall assume the masses of the scalars to be 
just as expected from the superpotential. The principle highlighted in the second of the
above papers concerns the almost automatic survival of $R$-parity in supersymmetric
left-right unification, and as a by-product, the almost pure Type I nature of the see-saw
mechanism. 

With above discussion in mind we pursue the models of  ~\cite{Aulakh:1998nn, Aulakh:1997ba} and \cite{Patra:2009wc}
to check the consistency of (i) gauge coupling unification and fermion mass universality, 
and (ii) the correct order of magnitude for the light neutrino masses, with the exciting 
possibility of (iii) TeV scale intermediate symmetry breaking. 
The issue of unification and perturbativity in this class of models  has recently been 
investigated exhaustively in \cite{Kopp:2009xt}. Our approach is similar in
spirit to the study of \cite{Dev:2009aw} for a different version of SUSYLR model and with different
motivations. Gauge coupling unification issue for the bitriplet Higgs model was studied
also in \cite{Borah:2010zq}. The present work extends this by the study of evolution of 
fermion masses and mixing for this model. A main finding of our paper, that at least for 
one of the models,  gauged $B-L$ charge can remain unbroken down to $3$ TeV can have 
interesting phenomenological consequences.

\indent This paper is organized as follows. In the next section \ref{Aulakh}, 
we discuss two versions of SUSYLR model ~\cite{Aulakh:1998nn, Aulakh:1997ba} as well as 
\cite{Patra:2009wc}. Then in section \ref{running} we study the gauge coupling unification 
and in section \ref{fermion} we study the evolution of fermion masses and mixing in both 
the models. We discuss the neutrino mass phenomenology in section 
\ref{neutrino} and then finally conclude in \ref{conclude}.

\section{Two possible choices for the SUSYLR Higgs structure}
\label{Aulakh}
The minimal set of the Higgs fields in the non-supersymmetric Left-Right model consists of a 
bidoublet $\Phi$ and $SU(2)_L$ and $SU(2)_R$ triplets $\Delta_L$ and $\Delta_R$ respectively. 
In the supersymmetric version, 
the cancellation of chiral anomalies among the fermionic partners of the triplet 
Higgs fields $\Delta$ requires introduction of the corresponding triplets $\bar{\Delta}$ 
with opposite $U(1)_{B-L}$ quantum number. Due to $B-L$ gauge invariance, the
$\Delta$ fields do not couple to the charged leptons and quarks, but gives majorana
masses to neutrinos upon getting a vev ( vacuum expectation value) while the
$\bar \Delta$ fields do not couple to fermions. The
usual fermion masses arise from a bidoublet $\Phi_u$. Another bidoublet $\Phi_d$ 
is introduced to avoid the Kobayashi-Maskawa matrix for quarks becoming trivial.
The matter supermultiplets of the minimal supersymmetric left-right model are
\begin{equation}
Q=
\left(\begin{array}{c}
\ u \\
\ d
\end{array}\right)
\sim (3,2,1,\frac{1}{3}),\hspace*{0.8cm}
Q_c=
\left(\begin{array}{c}
\ d_c \\
\ u_c
\end{array}\right)
\sim (3^*,1,2,-\frac{1}{3}),
\end{equation}
\begin{equation}
L=
\left(\begin{array}{c}
\ \nu \\
\ e
\end{array}\right)
\sim (1,2,1,-1), \quad
L_c=
\left(\begin{array}{c}
\ \nu_c \\
\ e_c
\end{array}\right)
\sim (1,1,2,1)
\end{equation}
where the numbers in the brackets denote the quantum numbers under 
$SU(3)_c \otimes SU(2)_L \otimes SU(2)_R \otimes U(1)_{B-L}$. 
Also here the convention is such that $L\rightarrow U_LL$ under 
$SU(2)_L$, but $L^c\rightarrow U^*_RL^c$ under $SU(2)_R$. 
The componentwise content of the scalar components of the Higgs superfields is as follows
\begin{equation}
\Phi_1=
\left(\begin{array}{cc}
\ \phi^0_{11} & \phi^+_{11} \\
\ \phi^-_{12} & \phi^0_{12}
\end{array}\right)
\sim (1,2,2,0),\hspace*{0.2cm} 
\Phi_2=
\left(\begin{array}{cc}
\ \phi^0_{21} & \phi^+_{21} \\
\ \phi^-_{22} & \phi^0_{22}
\end{array}\right)
\sim (1,2,2,0), 
\label{eq:Phicomponents}
\end{equation}
\begin{equation}
\bigtriangleup =
\left(\begin{array}{cc}
\ \frac{1}{\surd 2}\delta^+_L & \delta^{++}_L \\
\ \delta^0_L & - \frac{1}{\surd 2}\delta^+_L
\end{array}\right)
\sim (1,3,1,2), \hspace*{0.2cm}
\bar{\bigtriangleup} =
\left(\begin{array}{cc}
\ \frac{1}{\surd 2}\triangle^-_L & \triangle^0_L \\
\ \triangle^{--}_L & - \frac{1}{\surd 2}\triangle^-_L
\end{array}\right)
\sim (1,3,1,-2),
\label{eq:deltacomponents}
\end{equation}
\begin{equation}
\bigtriangleup_c =
\left(\begin{array}{cc}
\ \frac{1}{\surd 2}\triangle^-_R & \triangle^{--}_R \\
\ \triangle^0_R & - \frac{1}{\surd 2}\triangle^-_R
\end{array}\right)
\sim (1,1,3,-2), \hspace*{0.2cm}
\bar{\bigtriangleup}_c =
\left(\begin{array}{cc}
\ \frac{1}{\surd 2}\delta^+_R & \delta^0_R \\
\ \delta^{++}_R & - \frac{1}{\surd 2}\delta^+_R
\end{array}\right)
\sim (1,1,3,2) 
\label{eq:deltaCcomponents}
\end{equation}
Under left-right symmetry the fields transform as
\begin{equation}
Q\leftrightarrow Q^*_c,\quad L\leftrightarrow L^*_c,\quad \Phi\leftrightarrow 
\Phi^{\dagger},\quad \bigtriangleup \leftrightarrow \bigtriangleup^*_c,\quad  
\bar{\bigtriangleup}\leftrightarrow \bar{\bigtriangleup^*}_c  
\end{equation}

It turns out that left-right symmetry imposes rather strong constraints on
the ground state of this model. It was pointed out by Kuchimanchi and
Mohapatra \cite{Kuchimanchi:1993jg} that there is no spontaneous parity
breaking for this minimal choice of Higgs in the supersymmetric left-right 
model and as such the ground state remains parity symmetric. If parity odd 
singlets are  introduced to break this symmetry \cite{Cvetic:1985zp}, then it was
shown \cite{Kuchimanchi:1993jg} that the charge-breaking vacua have a
lower potential than the charge-preserving vacua and as such the ground
state does not conserve electric charge. Breaking $R$ parity was another
possible solution \cite{Kuchimanchi:1993jg} to this dilemma of breaking parity symmetry. 
For instance, it was shown recently in \cite{FileviezPerez:2008sx} that both Left-Right 
symmetry and R-parity can be broken simultaneously by right handed sneutrino vev.

A solution to this impasse without breaking $R$ parity is to add two 
new triplet superfields
$\Omega(1,3,1,0)$, $\Omega_c (1,1,3,0)$ where under parity symmetry
$\Omega \leftrightarrow \Omega_c^*$. This possibility has been explored
extensively in \cite{Aulakh:1997ba, Aulakh:1997fq, Aulakh:1998nn, Aulakh:1997vc},
which we refer to as the Aulakh-Bajc-Melfo-Rasin-Senjanovic (ABMRS) model. Another possibility is to add a Higgs bitriplet $\eta (1,3,3,0)$ and a parity odd singlet $\rho (1,1,1,0) $ \cite{Patra:2009wc} which also breaks parity spontaneously keeping R-parity conserved. We call this simply \textit{the bitriplet model}. We discuss both these models below.

\subsection{The ABMRS model}
As shown in the paper~\cite{Aulakh:1997ba}, the superpotential for this model is given by
\begin{eqnarray}
\lefteqn{W=h^{(i)}_l L^T\tau_2 \Phi_i \tau_2 L_c+ h^{(i)}_q Q^T\tau_2 
\Phi_i \tau_2 Q_c+ 
i fL^T\tau_2 \bigtriangleup L +
i f^*L^T_c 
\tau_2 \bigtriangleup_c L_c} \nonumber \\
&& +m_\triangle \text{Tr}\bigtriangleup \bar{\bigtriangleup}
+m^*_\Delta \text{Tr}\bigtriangleup_c \bar{\bigtriangleup}_c 
+\frac{m_\Omega}{2}\text{Tr}\Omega^2 +\frac{m^*_\Omega}{2} 
\text{Tr}\Omega^2_c \nonumber \\
&& +\mu_{ij}\text{Tr}\tau_2\Phi^T_i\tau_2\Phi_j +a\text{Tr}
\bigtriangleup \Omega \bar{\bigtriangleup}+a^*\text{Tr}
\bigtriangleup_c\Omega_c \bar{\bigtriangleup}_c \nonumber \\
&& +\alpha_{ij}\text{Tr}\Omega \Phi_i\tau_2 \Phi^T_j \tau _2 + 
\alpha^*_{ij}\text{Tr}\Omega_c \Phi^T_i\tau_2 \Phi_j \tau_2
\end{eqnarray}
where $h^{(i)}_{q,l}=h^{(i)\dagger}_{q,l},\mu_{ij}=\mu_{ji}=
\mu^*_{ij},\alpha_{ij}=-\alpha_{ji}$ and $f,h$ are symmetric matrices.
It is clear from the 
above superpotential that the theory has no baryon or lepton number violating 
terms. The Higgs fields either have $B-L=2$ or $0$. As such the spontaneous symmetry  
breaking automatically preserves $R$-parity defined by $(-1)^{3(B-L)+2S}$
.
Denoting the vev's of the neutral components of Higgs fields $\Omega_c$
and $\triangle_c$ to be $\omega_R$ and $v_R$, 
the supersymmetric vacua obtained  from the F-flatness conditions give the relationships 

\begin{equation}
\omega_R = -\frac{m_{\triangle}}{a}\equiv -M_R, \quad v_R= \sqrt{ \frac{2m_{\triangle}m_{\Omega}}{a^2}} \equiv M_{B-L} 
\label{mdeltamomega}
\end{equation}
The structure of $\Omega$ vev gives $SU(2)_R\rightarrow U(1)_R$. Thus if $v_R<\omega_R$ then the electroweak $U(1)_Y$
results only after the $\Delta$ fields get vev. 
The resulting symmetry breaking sequence in this case is 
\begin{displaymath}
SU(2)_L\otimes SU(2)_R\otimes U(1)_{B-L}\underrightarrow{\langle 
\Omega_c \rangle} SU(2)_L\otimes 
U(1)_R\otimes U(1)_{B-L}\underrightarrow{\langle \bigtriangleup_c \rangle} SU(2)_L\otimes U(1)_Y
\end{displaymath}
The choice $v_R<\omega_R$  also provides unambiguous parity breaking
from an early stage.
The two scales in question can  be kept distinct, by ensuring 
$M_{B-L} \ll M_R $ which can be achieved by choosing
$m_{\triangle} \gg m_{\Omega}$. 
A possibility for avoiding proliferation of new mass scales is to assume $m_\Omega=0$ in the
superpotential, by proposing an R-symmetry \cite{Aulakh:1997ba}. Then the physical value of 
$m_{\Omega}$ in the above relations can be assumed to arise from soft supersymmetry breaking terms, 
and hence of $M_{EW}$ scale. This then implies the relation $ M^2_{B-L} = M_{EW} M_R $ which relates 
different symmetry breaking scales.

In \cite{Aulakh:1997ba}, the vacumm structure was analysed
from $F$ flatness conditions.
For our purpose we consider the full scalar potential for the model given by 
\begin{equation}
V= \lvert F \rvert ^2 + D^a D^a/2 + V_{soft}
\end{equation}
where $F = \frac{\partial W}{\partial \phi} $ ,$D^a = -g(\phi^*_i T^a_{ij} \phi_j)$, g is gauge 
coupling constants, $T^a$ is the generators of the corresponding gauge group and $\phi$'s are 
chiral superfields, and $V_{soft}$ denotes all the soft supersymmetry breaking terms. We denote 
$\langle \triangle \rangle = v_L$, $\langle \triangle_c \rangle = v_R$, 
$\langle \Omega \rangle = \omega_L$, $\langle \Omega_c \rangle = \omega_R$, $\langle \Phi \rangle = v$. 
The soft terms can be ignored for pursuing the high scale physics, and the minimization of the 
scalar potential terms $V_F+V_D$ with respect to $v_L, v_R, \omega_L, \omega_R $ and $v$ respectively gives
\begin{equation}
v_L (m^2_{\triangle} + a \omega_L m_{\triangle} +a^2 \omega^2_L +m_{\Omega}\omega_L +av^2_L+ \alpha v^2+g^2 v^2_L) = 0
\label{eq1}
\end{equation}
\begin{equation}
v_R (m^2_{\triangle} - a \omega_R m_{\triangle} +a^2 \omega^2_R -m_{\Omega}\omega_R +av^2_R+ \alpha v^2+g^2 v^2_R) = 0
\label{eq2}
\end{equation}
\begin{equation}
a^2\omega_L v^2_L+a m_{\triangle} v^2_L + \omega_L m^2_{\Omega} +m_{\Omega}a v^2_L +\alpha v^2 m_{\Omega} + v^2 \alpha^2 \omega_R +g^2 \omega^3_L = 0
\label{eq3}
\end{equation}
\begin{equation}
a^2 \omega_R v^2_R-a m_{\triangle} v^2_R + \omega_R m^2_{\Omega} -m_{\Omega}a v^2_R +\alpha v^2 m_{\Omega} + v^2 \alpha^2 \omega_R +g^2 \omega^3_R = 0
\label{eq4}
\end{equation}
\begin{equation}
v \alpha (m_{\Omega} \omega_L+av^2_L+\omega_R m_{\Omega}-av^2_R+ \mu(\omega_L +\omega_R) +(\omega_L+\omega_R )^2 \alpha ) +g^2 v^3 = 0
\label{eq5}
\end{equation}
Note that it is sufficient for phenomenology to choose $v_R \neq 0$ 
and then it is natural to set $v_L=0$. In the non-supersymmetric version,
$v_L$ gets mixed with $v_R$ at tree level, and its value though small is 
not negligible
for the purpose of neutrino masses. This possibility is precluded here by the 
restriction imposed by supersymmetry. In turn this ensures pure Type I see-saw 
for the neutrino mass, assuming loop corrections to $v_L$ remain small. 
This also goes well with the requirements
$ v_L, \omega_L \ll M_{EW} $ so as not to affect the Standard Model 
$\rho$ parameter.

Our purpose here is to study the possibility of a TeV scale $U(1)_{B-L}$ breaking scale 
and a high $SU(2)_R$ breaking scale which can give rise to gauge coupling unification as 
well as small neutrino mass from seesaw.
Thus we choose $v_R \sim \text{TeV}, \omega_R \sim M_G $, where $M_G$ 
is the scale where the couplings unify. This can be ensured by taking 
$m_{\Omega} \sim $ TeV and  $m_{\triangle} \sim M_G$, with $M_G$
is expected to be $M_R$ introduced in Eq. (\ref{mdeltamomega}). 
With this choice, in the next section we study how the gauge couplings as 
well as fermion masses and mixings evolve under the renormalization group equations (RGE). 

\subsection{The Bitriplet Model}
The model above utilised  two mutually unrelated superfields $\Omega$
and $\Omega_c$.
We may attempt to achieve the same effect by invoking a bitriplet
superfield $\eta (1,3,3,0)$. This while separating the $M_R$ and the
$M_{B-L}$ scales as before, does not however succeed in providing
spontaneous parity breaking.
We are then led to add a parity odd singlet $\rho (1,1,1,0)$ 
to the particle content of minimal SUSYLR model  \cite{Patra:2009wc}. 
The superpotential with this Higgs content is 
\begin{eqnarray}
\lefteqn{W = f \eta_{\alpha i} \triangle_{\alpha} \triangle^c_i+f^* \eta_{\alpha i} 
\bar{\triangle}_{\alpha}\bar{\triangle}^c_i+ \lambda_1 \eta_{\alpha i} \Phi_{am} 
\Phi_{bn} (\tau^{\alpha} \epsilon)_{ab} (\tau^{i} \epsilon)_{mn}+m_{\eta} 
\eta_{\alpha i}\eta_{\alpha i}} \nonumber \\
&& +M_{\triangle}(\triangle_{\alpha}\bar{\triangle}_{\alpha}+\triangle^c_i 
\bar{\triangle}^c_i) +\mu \epsilon_{ab}\Phi_{bm}\epsilon_{mn}\Phi_{an}+ m_{\rho} 
\rho^2 +\lambda_2 \rho (\triangle_{\alpha}\bar{\triangle}_{\alpha}-\triangle^c_i 
\bar{\triangle}^c_i) 
\end{eqnarray}
where $\alpha, a, b$ are $SU(2)_L$ and $i, m, n$ are $SU(2)_R$     indices. The symmetry breaking pattern in this model is 
$$ SU(2)_L\times SU(2)_R\times U(1)_{B-L} \times P \quad \underrightarrow{\langle \eta \rangle} \quad SU(2)_L\times SU(2)_R\times U(1)_{B-L} $$
$$ \underrightarrow{\langle \bigtriangleup_c \rangle} \quad SU(2)_L\times U(1)_Y \quad \underrightarrow{\langle \Phi \rangle} \quad U(1)_{em} $$
Denoting the 
vev's as $\langle \triangle_- \rangle = \langle \bar{\triangle}_+ \rangle = v_L, 
\langle \triangle^c_+ \rangle = \langle \bar{\triangle}^c_- \rangle = v_R, \langle 
\Phi_{+-} \rangle = v, \langle \Phi_{-+} \rangle = v', \langle \eta_{+-} \rangle = 
u_1, \langle \eta_{-+} \rangle = u_2, \langle \eta_{00} \rangle = u_0 $. 
The scalar potential is $V = V_F+V_D+V_{soft}$ where $V_F = \lvert F_i \rvert^2, F_i = 
-\frac{\partial W}{\partial \phi} $ is the F-term scalar potential, $V_D = D^a D^a/2, D^a = 
-g(\phi^*_i T^a_{ij} \phi_j)$ is the D-term 
of the scalar potential and $ V_{soft}$ is the soft supersymmetry breaking terms in the 
scalar potential. Ignoring the soft terms as before for analysis of the high scale physics we have 
\begin{eqnarray}
\lefteqn{\frac{\partial V}{\partial v_L} = \mu^2_L (2v_L) +2 \lambda^2_2 v_L(v^2_L-v^2_R) 
+2(fu_1+f^* u_2)M_{\triangle} v_R } \nonumber \\
&& +v_R(f+f^*)[2m_{\eta} (u_1+ u_2+u_3) +\lambda_1 v^2 +v_Lv_R(f+f^*)] = 0
\label{eq10}
\end{eqnarray}
\begin{eqnarray}
\lefteqn{\frac{\partial V}{\partial v_R} = \mu^2_R (2v_R) -2 \lambda^2_2 v_R(v^2_L-v^2_R) 
+2(fu_1+f^* u_2)M_{\triangle} v_L } \nonumber \\
&& +v_L(f+f^*)[2m_{\eta} (u_1+ u_2+u_3) +\lambda_1 v^2 +v_Lv_R(f+f^*)] = 0
\label{eq11}
\end{eqnarray}
Where the effective mass terms $\mu^2_L, \mu^2_R $ are given by
\be
\mu^2_L = (M_{\triangle}+\lambda_2 s)^2 + \lambda_2 m_{\rho}s +\frac{1}{2}(f^2u^2_1+f^{*2}u^2_2)
\ee
\be
\mu^2_R = (M_{\triangle}-\lambda_2 s)^2 -\lambda_2 m_{\rho}s +\frac{1}{2}(f^2u^2_1+f^{*2}u^2_2)
\ee
Thus after the singlet field $\rho$ acquires a vev the degeneracy of the Higgs triplets goes 
away and the left handed triplets being very heavy get decoupled whereas the right handed 
triplets can be as light as 1 TeV by appropriate fine tuning in the above two expressions. 
Assuming $v_L \ll v_R \ll m_{\rho}, m_s$ we get from equations (\ref{eq10}), (\ref{eq11}):
\be
v_L = \frac{-v_R[M_{\triangle} u_2 f^* +m_{\eta} (u_2+u_3)(f+f^*)+u_1(fM_{\triangle}+ 
m_{\eta}(f+f^*)]}{2m_{\rho}s\lambda_2 +4 M_{\triangle}s\lambda_2}
\ee
To understand this relation let us assume $s\sim M_\triangle\sim m_\eta\sim m_\rho$ collectively
denoted by $M_R$ to be large, and $u_1\sim u_2\sim u_3$ denoted $u$ to be small. The above relation 
then reads, ignoring dimensionless numbers,
$$
v_L \approx v_R \times \frac{u}{M_R}
$$
We must take the vev of the bitriplet $u \ll M_Z$ so as not to affect the Standard Model 
$\rho$ parameter. On the other hand $v_L$ which enters the see-saw formula has maximally allowed
value $\sim$eV. 
Thus if we want $v_R$ to be low, $\sim 1 \text{TeV}$, possibly giving rise to
collider signatures, then the above relation when saturated requires the scale of parity 
breaking to be kept low compared to GUT scale, 
$M_R \sim 10^{10}$GeV. But we shall  see that gauge coupling unification forces $M_R$ 
to be much higher. This leaves a large 
parameter space  for the possible values of $v_L$ and 
$v_R$, such that they remain phenomenologically accessible. In particular, retaining 
Type I see-saw for neutrino masses remains natural.

\section{Gauge Coupling Unification}
\label{running}
The one-loop renormalization group evolution equations \cite{Jones:1981we} are given by 
\begin{equation}
\mu \frac{ d g_i}{d\mu} = \beta_i(g_i) = \frac{g^3_i}{16 \pi^2} b_i
\label{eq6}
\end{equation}
Defining $\alpha_i = g^2_i/(4\pi) $ and $ t = ln( \mu/\mu_0) $ and the most general renormalization group equation above becomes
\begin{equation}
\frac{d \alpha^{-1}_i}{dt} = -\frac{b_i}{2 \pi}
\label{eq7}
\end{equation}
The one-loop beta function is given by
\begin{equation}
\beta_i(g_i)=\frac{g^3_i}{16 \pi^2}[-\frac{11}{3}\text{Tr}[T^2_a]+\frac{2}{3}\sum_{f} \text{Tr}[T^2_f]+\frac{1}{3}\sum_{s} \text{Tr}[T^2_s]]
\end{equation}
where $f$ means the fermions and $s$ means the scalars. For $SU(N)$, Tr$[T^2_a] =N$ and $Tr[T_i T_i]=\frac{1}{2}$. For a supersymmetric model the most general beta function is  given by
\begin{equation}
\beta_i(g_i)=\frac{g^3_i}{16 \pi^2}[-3\text{Tr}[T^2_a]+\sum_{f} \text{Tr}[T^2_f]+\sum_{s} \text{Tr}[T^2_s]]
\end{equation}
The one-loop renormalization group evolutions(RGE) for the masses in SUSYLR model have already been calculated analytically in \cite{Setzer:2005hg} whereas the same for Minimal Supersymmetric Standard Model (MSSM) can be found in \cite{Das:2000uk}. A very recent analysis on the evolution of fermion masses and mixing was carried out in \cite{Dev:2009aw}. We will use the analytical results from these references to study the gauge couplings, fermion mass and mixing evolution in both ABMRS and the bitriplet model below.

\subsection{The ABMRS model}
\label{subsection:ggeABMRS}
For the particle content of the ABMRS model we calculate the beta functions as follows
\begin{itemize}
\item Below the SUSY breaking scale $M_{susy}$ the beta functions are same as those of the standard model
$$ b_s = -11+\frac{4}{3}n_g,\quad b_{2L} = -\frac{22}{3}+\frac{4}{3}n_g+\frac{1}{6}n_b, \quad b_Y = \frac{4}{3} n_g + \frac{n_b}{10} $$
\item For $ M_{susy} < M < M_{B-L} $ , the beta functions are same as those of the MSSM
$$ b_s = -9+2n_g,\quad b_{2L} = -6+2n_g+\frac{n_b}{2}, \quad  b_Y = 2n_g+\frac{3}{10}n_b$$
\item For $ M_{B-L} < M < M_R $ the beta funtions are 
$$ b_s = -9+2n_g,\quad  b_{2L} = -6 +2n_g +\frac{n_b}{2}+2 n_{\Omega} $$
$$ b_{1R} = 2n_g + \frac{n_b}{2} + 2 n_{\Omega},\quad b_{B-L} = 2n_g $$
\item For $ M_R < M < M_{GUT} $ the beta functions are 
$$ b_s = -9+2n_g,\quad b_{2L} = -6 +2n_g +n_b+ 2 n_{\Omega} +2 n_{\triangle} $$
$$ b_{2R} = -6+2n_g + n_b + 2 n_{\Omega} + 2 n_{\triangle},\quad b_{B-L} = 2n_g + 9 n_{\triangle} $$
\end{itemize}
Where $n_g = 3, n_b = 2, n_{\Omega} = 1, n_{\triangle} = 2$ are the number of generations , number of bidoublets, number of $\Omega$ and number of $\triangle$ respectively. 

\begin{figure}[ht]
 \centering
\includegraphics{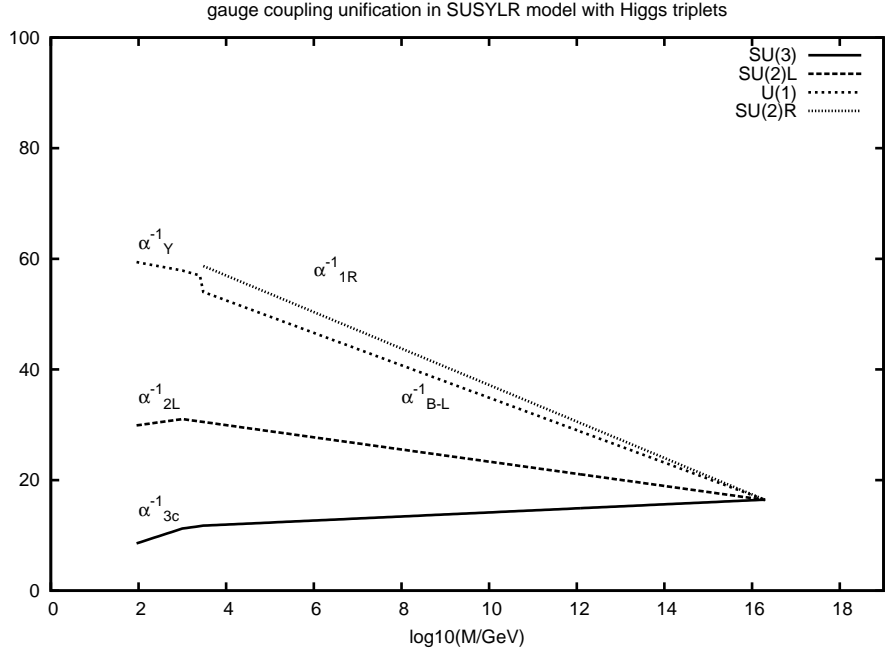}
\caption{Gauge coupling unification in the ABMRS model with two extra pairs of colored superfields $\chi_{1,2}(3,1,1,-\frac{2}{3}), \bar{\chi}_{1,2}(\bar{3},1,1,\frac{2}{3})$, $M_{SUSY} = 1$ TeV,$m_{\Omega}=M_{B-L} = 3$ TeV, $m_{\triangle}=M_{R} =M_{\text{GUT}}= 2 \times 10^{16}$ GeV. The extra superfields decouple below $M_{B-L}$.}
 \label{fig1}
\end{figure}

It is found that with just the particle content of the SUSYLR model discussed above, the gauge couplings do not 
unify because of  too fast a running of the coupling $\alpha_{3c}$. Additional
colored superfields are needed to achieve unification. We find that two pairs of extra 
superfields $\chi_{1,2}(3,1,1,-\frac{2}{3}), \bar{\chi}_{1,2}(\bar{3},1,1,\frac{2}{3})$, singlet under 
the $SU(2)_{L,R}$ are needed for the gauge couplings to unify. 
Each of these extra superfields contributes $\frac{1}{2}$ to both the beta functions 
$b_s$ and $b_{B-L}$ and does
not affect the other beta functions, while the vectorlike choice of charges ensures no anomalies. 
Contributions of a variety of such new superfields to 
the beta functions were calculated in \cite{Kopp:2009xt} and our result is in agreement with them. Interestingly,
while this many additional fields are just sufficient to achieve the required goal, any additional
added fields will drive the coupling into the Landau pole. The resulting unification is shown in figure \ref{fig1}. 
These extra superfields can naturally be accommodated within $SO(10)$ GUT representations, 
either $\textbf{120}$ or $\overline{\textbf{126}}$. Here we assume that masses of these extra superfields can be as low as the $U(1)_{B-L}$ breaking scale.

\begin{figure}[ht]
\centering
\includegraphics{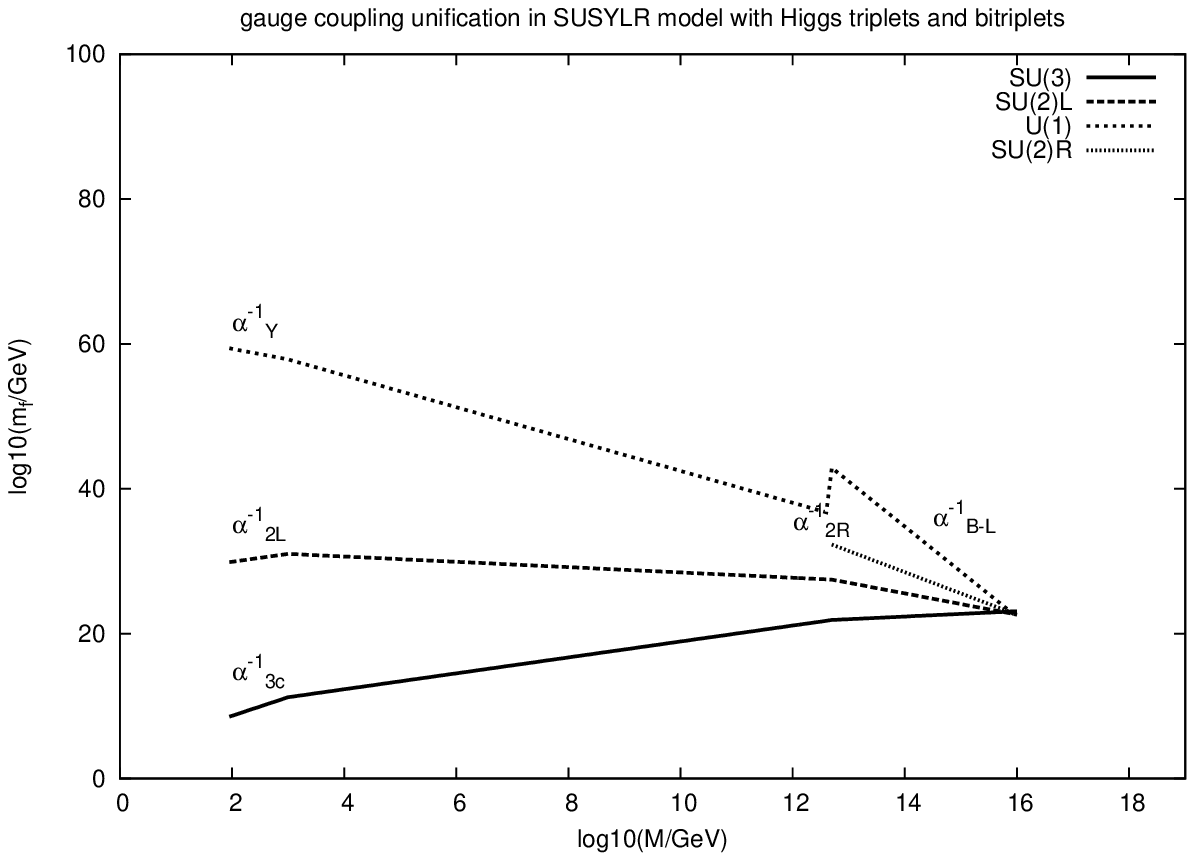}
\caption{Gauge coupling unification in the bitriplet model with two extra pairs of colored 
superfields $\chi_{1,2}(3,1,1,-\frac{2}{3}), \bar{\chi}_{1,2}(\bar{3},1,1,\frac{2}{3})$, $M_{susy} = 1$ TeV, $M_R = 5\times 10^{12} $ GeV, $M_{\text{GUT}}=M_{\rho} = 10^{16}$ GeV. The extra superfields decouple below the scale $M_R$.}
\label{fig2}
\end{figure}
\subsection{The Bitriplet Model}
It is found that the gauge couplings do not unify for the minimal field content 
of the bitriplet model.  Similar to the AMBRS model, it is adequate to add four heavy colored 
superfields $\chi_{1,2}(3,1,1,-\frac{2}{3})$, $\bar{\chi}_{1,2}(\bar{3},1,1,\frac{2}{3}) $.
And these are then required to decouple below the $SU(2)_R $ breaking scale $M_R$. 
The beta functions above $ M_R$ are 
\begin{itemize}
\item For $M_R < M < M_{\rho}$ the beta functions are
$$ b_s = -9+2n_g + \frac{n_{\chi}}{2},\quad b_{2L} = -6 +2n_g +n_b + 2n_{\eta} $$
$$ b_{2R} = -6+2n_g + n_b+2n_{\triangle} + 2n_{\eta},\quad b_{B-L} = 2n_g +\frac{9}{2}n_{\triangle}+\frac{n_{\chi}}{2}$$
\item For $ \langle \rho \rangle < M < M_{GUT} $ the beta functions are 
$$ b_s = -9+2n_g+  \frac{n_{\chi}}{2},\quad b_{2L} = -6 +2n_g +n_b+2 n_{\triangle}+2 n_{\eta} $$
$$ b_{2R} = -6+2n_g + n_b + 2 n_{\triangle}+ 2 n_{\eta},\quad b_{B-L} = 2n_g + 9n_{\triangle}+\frac{n_{\chi}}{2} $$
\end{itemize}
where $n_{\triangle} = 2, n_{\chi}=4 , n_g = 3, n_b = 2, n_{\eta}=1 $. 
Using the same initial values and normalization relations as before 
we arrive at the gauge coupling unification, an essential result of \cite{Borah:2010zq}, as 
shown in Fig. \ref{fig2}.  Here the unification scale is the same as the 
$D$-parity breaking scale. 
Similar to the case with just Higgs triplets, here also lower value of 
$M_R$ makes the unification look worse. Thus although minimization of 
the scalar potential allows the possibility of a TeV scale $M_R$ in 
this model, the requirement of gauge coupling unification rules out such a 
possibility.

\section{Running Fermion Masses and Mixings}
\label{fermion}
Assuming that the gauge coupling unification is achieved due to the presence of
the additional colored multiplets as discussed in the previous section, we consider
the question of fermion mass universality and fermion mixing.  
We also do this in the same context as in the previous section by fixing the intermediate symmetry 
breaking scales to  those which gave rise to gauge coupling unification as in Fig. \ref{fig1} and Fig. \ref{fig2}.
To analyse the fermion mass running we consider all the leptonic yukawa couplings 
to be diagonal for simplicity. We take the initial values of the masses and mixing parameters 
at the electroweak scale 
from \cite{Amsler:2008zzb}. After fixing all these, we are still left with the freedom of choosing the 
couplings $f, f^*$ at the electroweak scale and the ratio of the two electroweak vevs $ \langle \Phi_1 \rangle = \text{diag}(v_1, 0), \langle \Phi_2 \rangle = \text{diag}(0, v_2)$  which we denote as $\tan{\beta} = \frac{v_1}{v_2}$. 
Within the context of a simple analysis, we assume $f, f^*$ to be diagonal, and proportional to the identity
matrix at the electroweak scale. 

\begin{figure}[htb]
 \centering
\includegraphics{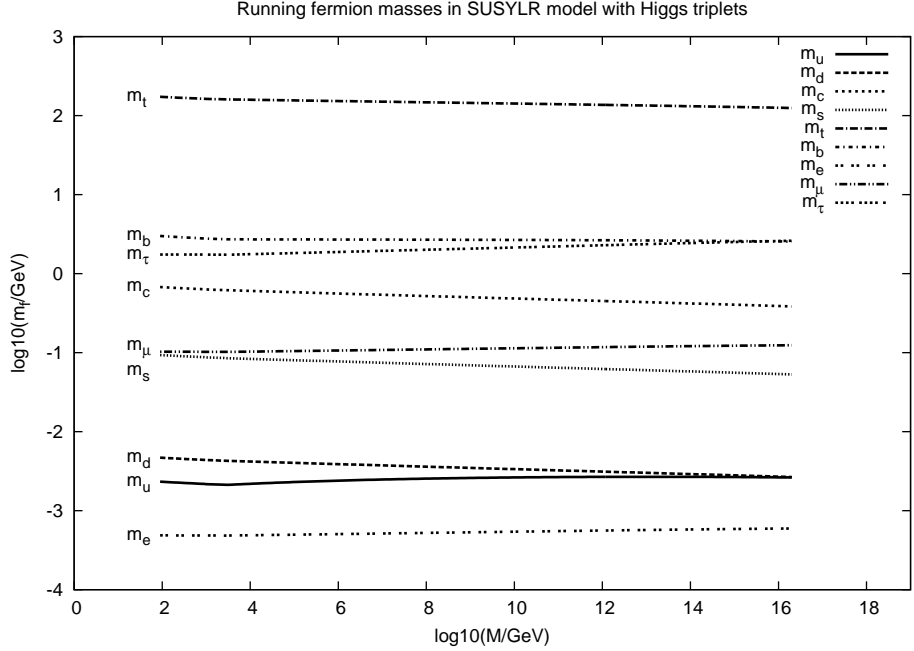}
\caption{Running fermion masses in the ABMRS model with  extra colored superfields $\chi_{1,2}(3,1,1,-\frac{2}{3})$, $\bar{\chi}_{1,2}(\bar{3},1,1,\frac{2}{3})$, $M_{SUSY} = 1$ TeV, $m_{\Omega}=M_{B-L} = 3$ TeV, $m_{\triangle}=M_{R}=M_{\text{GUT}} = 2 \times 10^{16}$ GeV and $|f|=0.55, \tan{\beta} = 10 $ at $M=M_Z$}
 \label{fig3}
\end{figure}

\begin{figure}[htb]
\centering
\includegraphics{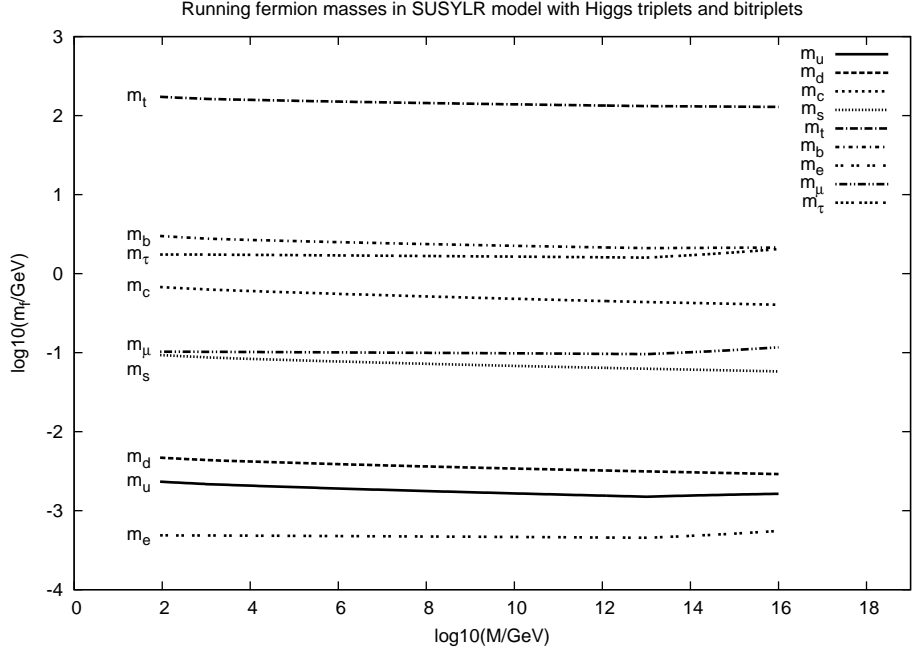}
\caption{Fermion masses evolution in the bitriplet model with extra colored superfields $\chi_{1,2}(3,1,1,-\frac{2}{3})$, $\bar{\chi}_{1,2}(\bar{3},1,1,\frac{2}{3})$, $M_{susy} = 1$ TeV, $M_R = 5 \times 10^{12} $ GeV, $M_{\rho} = 10^{16}$ GeV and $|f| = 0.90, \tan{\beta} = 10$ at $M = M_Z$}
\label{fig4}
\end{figure}

Before presenting the general answer for the allowed range of 
parameters $|f|$ and $\tan\beta$ consistent with $b-\tau$ unification, let us consider 
specific successful pairs of values which work for each of the models.
The predictions for fermion masses and mixings  at the 
GUT scale for ABMRS model, with $|f|=0.55$ and $\tan\beta=10$, and for the bitriplet model
with $|f|=0.90$ and $\tan\beta=10$ are shown in  table \ref{table1}. 
The running of fermion masses in detail 
are shown in Fig. \ref{fig3} for the ABMRS model and in Fig. \ref{fig4} for the bitriplet
model.
\begin{center}
\begin{table}[ht]
\caption{Running Fermion masses in SUSYLR model for $\tan{\beta}=10$}
\begin{tabular}{|c|c|c|c|}
\hline
Fermion Masses      & $M=M_Z$    & $M=M_G$ (ABMRS)  &  $M=M_G$ (Bitriplet) \\ 
          & PDG \cite{Amsler:2008zzb}     & ($|f| = 0.55$) & ($|f| = 0.90)$  \\
\hline
$m_u $(MeV)      &  $2.33^{+0.42}_{-0.45}$      & $2.629$   & $1.635$      \\
$m_d$(MeV)    &  $4.69^{+0.60}_{-0.66}$       & $ 2.659$& $ 2.905$                          \\ 
$m_c$(MeV)        &  $677^{+56}_{-61}$   &  $383.654$      &  $403.89$           \\ 
$m_s$(MeV)           &  $93.4^{+11.8}_{-13.0}$      & $52.924$ & $57.881$                       \\
$m_t$(GeV)        &  $181\pm 13$      &$124.511$    &$128.69$              \\
$m_b$(GeV)           &  $3.0\pm 0.11$      & $2.542$& $2.138$                       \\
$m_e$(MeV)      & $0.48684727 \pm 0.14 \times 10^{-6}$    & $0.5953$   & $0.5549$ \\
$m_{\mu}$(MeV)      & $102.75138\pm 3.3\times 10^{-4}$  & $124.22$ & $116.823$   \\
$m_{\tau}$(GeV)      & $1.74669^{+0.00030}_{-0.00027} $  & $2.615$   & $2.046$  \\
\hline
\end{tabular}
\label{table1}
\end{table}
\end{center}
\begin{center}
\begin{table}[t]
\caption{Running CKM elements in SUSYLR model for $\tan{\beta}=10$}
\begin{tabular}{|c|c|c|c|}
\hline
CKM elements      & $M=M_Z$    & $M=M_G$ (ABMRS) & $M=M_G$ (Bitriplet)  \\ 
          &PDG \cite{Amsler:2008zzb}      & ($|f| = 0.55$) & ($|f| = 0.90$)   \\
\hline
$V_{ud}$      &  $0.9742$      & $0.9780$    & $0.9777$  \\
$V_{us}$    &  $0.2256$       & $ 0.208-0.0008i$  & $ 0.210-0.0008i$                     \\ 
$V_{ub}$        &  $0.0013-0.0033i$   &  $0.0006-0.0021i$  & $0.0006-0.0021i$    \\ 
$V_{cd}$           &  $-0.2255-0.0001i$      & $-0.208-0.0009i$& $-0.210-0.0008i$                       \\
$V_{cs}$        &  $0.9734$      &$0.9777$        &$0.9773$      \\
$V_{cb}$           &  $0.0415$      & $0.0269$     & $0.0268$                  \\
$V_{td}$      & $0.0081-0.0032i$    & $0.0072-0.0029i$ & $0.0072-0.0029i$ \\
$V_{ts}$      & $-0.0407-0.0007i$  & $-0.0378-0.0006i$   & $-0.0378-0.0006i$ \\
$V_{tb}$      & $0.9991$  & $0.9851$     & $0.9850$   \\
\hline
\end{tabular}
\label{table2}
\end{table}
\end{center}
\begin{figure}[htb]
 \centering
\includegraphics{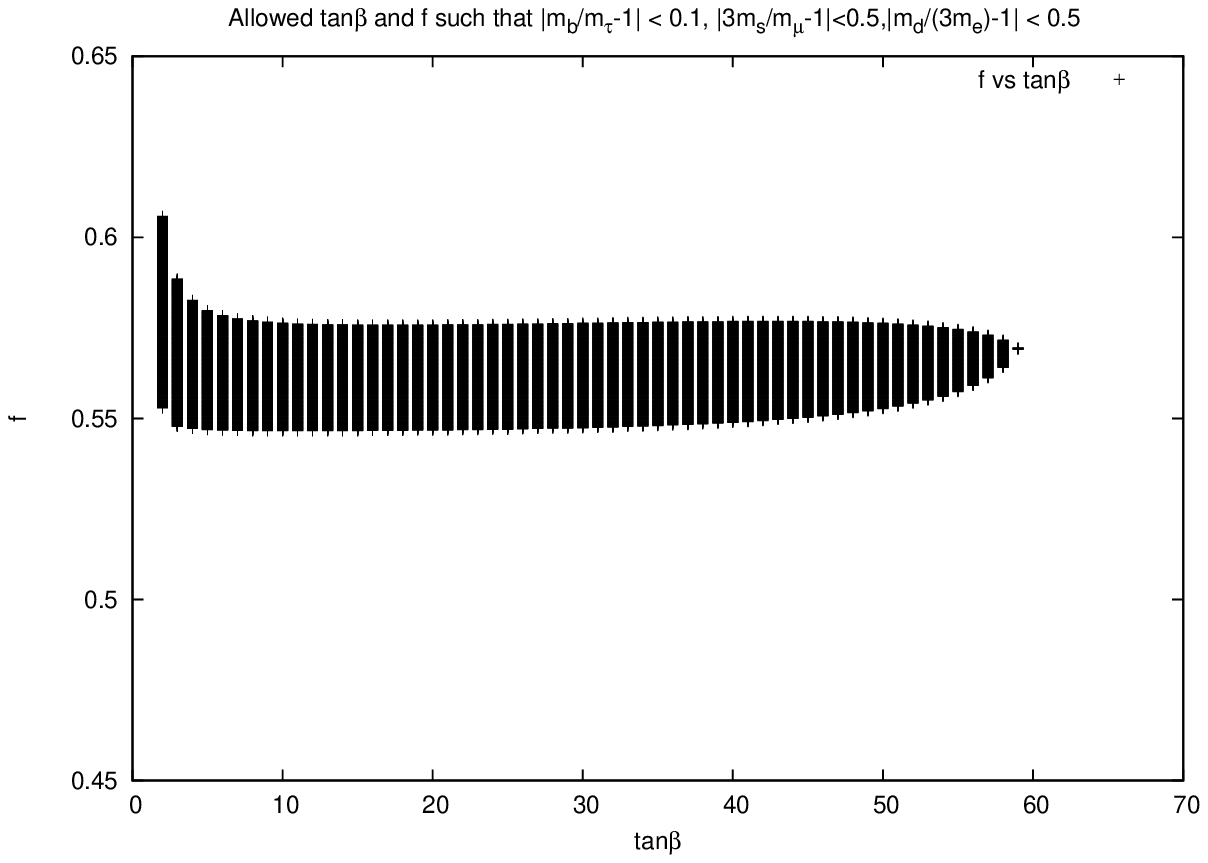}
\caption{Allowed parameter ranges of $f$ and $\tan{\beta}$ such that  $\lvert m_b/m_{\tau}(M_G)-1 \rvert < 0.1, \lvert \frac{3m_s}{m_{\mu}}(M_G)-1 \rvert < 0.5, \lvert \frac{m_d}{3m_e}(M_G)-1 \rvert < 0.5$  for ABMRS model with extra colored superfields $\chi_{1,2}(3,1,1,-\frac{2}{3}), \bar{\chi}_{1,2}(\bar{3},1,1,\frac{2}{3})$, $M_{SUSY} = 1$ TeV,$m_{\Omega}=M_{B-L} = 3$ TeV, $m_{\triangle}=M_{R} =M_{\text{GUT}}= 2 \times 10^{16}$ GeV}
 \label{fig5}
\end{figure}
\begin{figure}[htb]
 \centering
\includegraphics{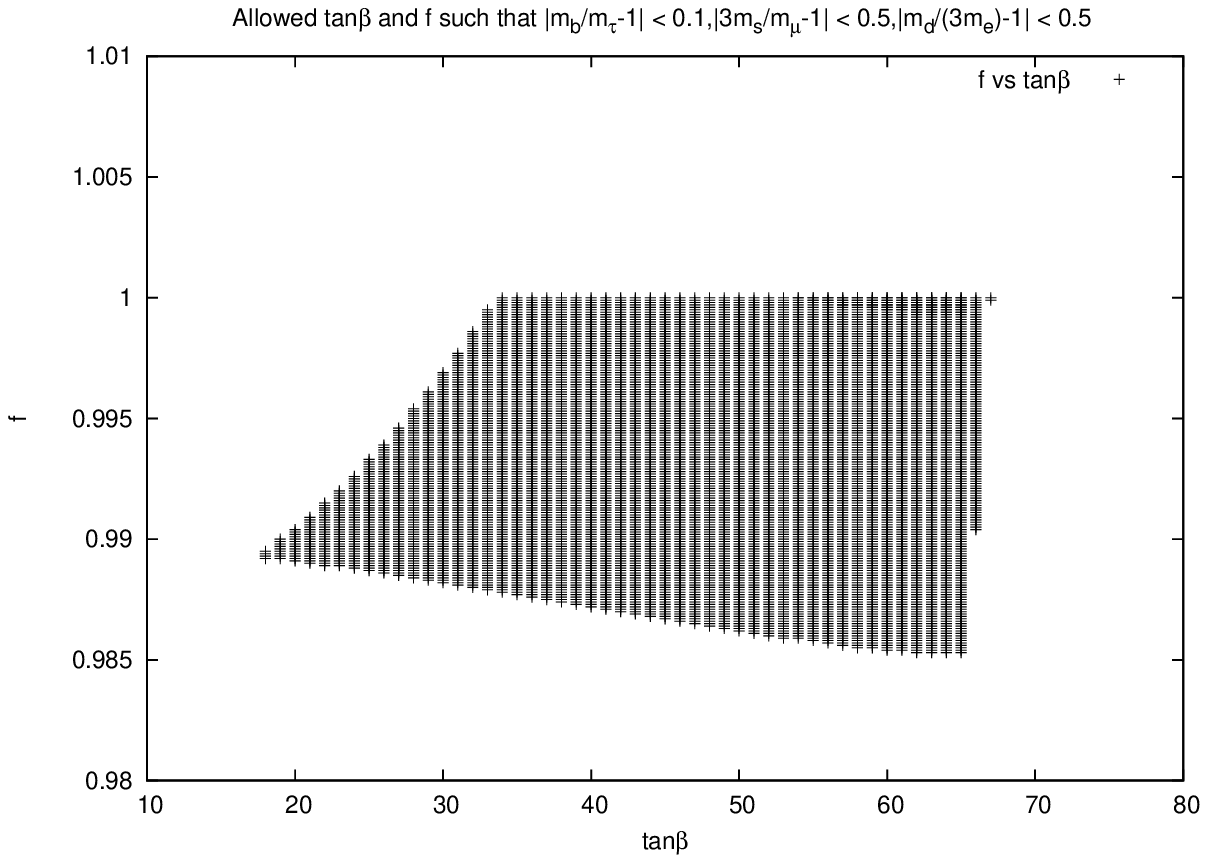}
\caption{Allowed parameter ranges of $f$ and $\tan{\beta}$ such that $ \lvert \frac{m_b}{m_{\tau}}(M_G)-1 \rvert < 0.1, \lvert \frac{3m_s}{m_{\mu}}(M_G)-1 \rvert < 0.5, \lvert \frac{m_d}{3m_e}(M_G)-1 \rvert < 0.5$ for the bitriplet model with extra colored superfields $\chi_{1,2}(3,1,1,-\frac{2}{3}), \bar{\chi}_{1,2}(\bar{3},1,1,\frac{2}{3})$, $M_{SUSY} = 1$ TeV, $M_{R} = 5 \times 10^{12}$ GeV and $M_{\rho}=M_{GUT} = 10^{16} $ GeV}
 \label{fig6}
\end{figure}

At the GUT scale the ratios of fermion masses come out to be
\begin{equation}
\frac{m_b}{m_{\tau}}=0.9720, \quad \frac{3m_{s}}{m_{\mu}} = 1.2781, \quad \frac{m_d}{3m_e} = 1.4888 
\end{equation}
for the ABMRS model, and 
\begin{equation}
\frac{m_b}{m_{\tau}}=1.0449, \quad \frac{3m_{s}}{m_{\mu}} = 1.4863, 
\quad \frac{m_d}{3m_e} = 1.7450 
\end{equation}
for the bitriplet model.
These ratios are expected to be unity in a grand unified theory \cite{Georgi:1979df}. The mismatch in the
values, especially that for the lighter quarks and leptons is expected to be corrected by the incorporation of various 
threshold effects \cite{Banks:1987iu,Hempfling:1993kv,Blazek:1995nv,DiazCruz:2000mn,Antusch:2008tf,Aulakh:2008sn,Enkhbat:2009jt}.  
We also study the running of Cabbibo Kobayashi Maskawa (CKM) elements and their predicted  values at the GUT scale are mentioned in table \ref{table2}. \\

\indent These results can be generalised by allowing a variation of  $m_b/m_{\tau}(M_G)$  
with respect to the initial choices of $\tan{\beta}$ and the yukawa 
coupling $|f|$.
We plot the allowed range of parameters 
$(|f|,\tan{\beta})$ for which $ \lvert m_b/m_{\tau}(M_G)-1 \rvert < 0.1, 
\lvert \frac{3m_s}{m_{\mu}}(M_G)-1 \rvert < 0.5, 
\lvert \frac{m_d}{3m_e}(M_G)-1 \rvert < 0.5 $. 
Usually there is far less discrepancy in the case of third generation 
fermion universality at the unification scale, and hence we allow
variation of $10\%$ error in its value, expecting the discrepancy to be remedied 
easily by incorporating various corrections. The discrepancy in case 
of lighter fermions are much more and can be removed only after 
considering radiative corrections \cite{Banks:1987iu,Hempfling:1993kv,Blazek:1995nv,DiazCruz:2000mn,Antusch:2008tf,Aulakh:2008sn,Enkhbat:2009jt}. 
We have retained a larger tolerance of $50\%$ in those ratios.
The corresponding plots are shown in Fig.s \ref{fig5} and  \ref{fig6}. 
It can be seen that only a narrow range of the yukawa coupling
$|f|$
at the electroweak scale leads to $b-\tau$ unification at 
the GUT scale whereas a wide range of $\tan{\beta}$ values can give rise to the same.

\section{Neutrino mass}
\label{neutrino}
\indent The type I and type II contributions to the generalised see-saw relation for the 
light neutrino mass matrix $m_{\nu ij} $ in the  left-right models are characterised respectively as 
\begin{equation}
m^{I}_{\nu ij} = -(M_D M^{-1}_R M^T_D)_{ij}; \qquad m^{II}_{\nu ij} = f_{ij} v_L  
\label{nu1}
\end{equation}
where $M_D$ is the dirac mass matrix of the neutrinos $(M_D)_{ij} = h_{ij} v_1$. 
In the ABMRS model, as discussed below Eq.s (\ref{eq1})-(\ref{eq5}) in sub-section \ref{Aulakh} 
$v_L = 0$ is a natural value for $\langle \triangle \rangle$, characteristic of the incorporation
of supersymmetry. Thus the first term in the neutrino mass formula (\ref{nu1}) vanishes and only 
the second term survives making type I seesaw natural in the ABMRS model.

\indent In the bitriplet model the above formula (\ref{nu1}) can be written as
\begin{equation}
m_\nu\equiv m^{II}_\nu+m^{I}_\nu =\frac{-f\,v^2\,v_R}{2\, m_\sigma\,s}-\frac{v^2}{v_R} h\, f^{-1}\,h^T
\end{equation}
Attempting to keep $v_R \sim 1$ TeV accessible to accelerator energies, the
Type II contribution to the small neutrino masses can be kept within observed limits 
provided $m_{\sigma} \sim s $ are at least $\gtrsim 10^8-10^{10}$ GeV. 
Here, as in \cite{Borah:2010zq}, we find that the values of these mass scales
should in fact be much higher, closer to GUT scale $M_G$ in oder to achieve
gauge coupling unification.  This renders the Type II contribution completely negligible 
even with a $v_R$ scale as high $\sim M_G$. The bitriplet model does introduce more scales than 
minimally required. However, the new scales introduced are stabilised by supersymmetry.
This leaves open the phenomenologically interesting possibility of choosing
a TeV scale $v_R$, which can potentially enter new physics signatures in collider data.

\section{Results and Conclusion}
\label{conclude}
Supersymmetric version of the Left-Right symmetric model including
automatic Majorana mass for the neutrinos is insufficient to
provide spontaneous parity breaking as a renormalizable theroy
unless the Higgs structure is suitably extended. 
We have considered two possible extensions, ABMRS model with additional
Higgs triplets and another model with an additional bitriplet as well 
as a parity odd singlet. In each of these we have studied the 
evolution of gauge couplings and fermion masses
and mixings. In the ABMRS model, a particular choice of scales 
$M_{B-L} \sim \text{TeV}, 
m_{\Omega} \sim \text{TeV}, m_{\triangle} \sim M_R \sim M_G $  
is identified, giving rise to gauge coupling as well as $b-\tau$ 
unification, but which demands inclusion of two pairs of additional heavy colored superfields 
$\chi_{1,2}(3,1,1,-\frac{2}{3}), \bar{\chi}_{1,2}(\bar{3},1,1,\frac{2}{3})$. 
Similar analysis in the case of the bitriplet 
model also requires two additional pairs of heavy colored 
superfields $\chi_{1,2}(3,1,1,-\frac{2}{3}), \bar{\chi}_{1,2}(\bar{3},1,1,\frac{2}{3})$,
 which in this case decouple below the scale $M_R$. 
These extra superfields can be naturally embedded within $SO(10)$ GUT 
representations, $\textbf{120}$ or $\overline{\textbf{126}}$.

In the ABMRS case, it is possible to have a really low
$M_{B-L}$ of 3 TeV, though the scale $M_R$ is required to be close to the Grand Unified Theory (GUT) scale. 
This has several interesting phenomenologically testable
consequences. Firstly it makes it possible to explore some aspects 
of the breaking of the $B-L$ quantum number at collider energies.
For example, the scale of the majorana masses of the neutrino would also
be at this scale and make the physics of lepton number violation
accessible to colliders. Also, this means that any baryon asymmetry 
of the Universe to be generated should have occurred only at a scale 
lower than this low scale. It has been shown that majorana masses at
this scale and lighter do not conflict with baryogenesis via leptogenesis 
provided  leptogenesis itself is non-thermal \cite{Sahu:2004sb}.
Specifically, the $CP$ violation  mechanism  involved in the creation 
of asymmetry becomes accessible to collider energies.

In the bitriplet model there appear to be several
adjustable energy scales. The gauge coupling unification necessarily 
forces the $SU(2)_R$ breaking to be at a high scale $M_R \geq 5 \times 10^{12}$GeV. 
However supersymmetry may yet protect new scales much smaller
than this scale, and allow a much lower value of $v_R$. Existence
of such new scales may provide interesting windows to new physics 
accessible to accelerators.

We have also identified the parameter space at the electroweak scale 
which gives rise to fermion mass universality at the unification 
scale for both the ABMRS and the bitriplet model. 

Finally, we see that tiny neutrino mass arises from the type I seesaw 
in the ABMRS model as earlier proposed. Further, in the bitriplet model,
despite the non-zero Type II contribution at tree level, the latter
contribution is rendered utterly negligible due to the required high
scale of gauge coupling unification. Thus Type I see-saw emerges
as the natural explanation of the small neutrino masses in both the models.

\section{Acknowledgement} 
We thank Charanjit Aulakh for critical comments. This work is a part of a project 
supported by a grant from DST, India. UAY thanks the members of the 
McGill High Energy Theory group for hospitality and financial support during 
a sabbatical visit.
DB would like to acknowledge the hospitality of Physical Research 
Laboratory, Ahmedabad where most of the work was done.  
He is also grateful to IIT Bombay for giving him a leave of absence 
to spend one semester at Physical Research Laboratory, Ahmedabad.


\end{document}